\documentclass[conference]{IEEEtran}
\IEEEoverridecommandlockouts
\usepackage{cite}
\usepackage{amsmath,amssymb,amsfonts}
\usepackage{algorithmic}
\usepackage{graphicx}
\usepackage{textcomp}
\usepackage{xcolor}
\usepackage{multirow} 
\usepackage{gensymb} 
\usepackage{comment} 
\def\BibTeX{{\rm B\kern-.05em{\sc i\kern-.025em b}\kern-.08em
    T\kern-.1667em\lower.7ex\hbox{E}\kern-.125emX}}
    \usepackage{academicons}
\definecolor{orcidlogocol}{HTML}{A6CE39}
\begin{document}

\title{A Performance Investigation of Receive Beamforming Schemes in Specular Tissue Characterization\\
\thanks{This work is supported by Department of Science and Technology - Science and Engineering Research Board (DST-SERB ECR/2018/001746) and the Ministry of Education, India.}
}
\author{\IEEEauthorblockN{Gayathri Malamal and Mahesh Raveendranatha Panicker}
\IEEEauthorblockA{\textit{Center for Computational Imaging, Dept. of Electrical Engineering} \\
\textit{Indian Institute of Technology Palakkad, Kerala, India}\\
121814001@smail.iitpkd.ac.in;mahesh@iitpkd.ac.in}
}

\maketitle

\begin{abstract}
This work presents a comparison of the delay and sum (DAS), filtered delay multiply and sum (DMAS), minimum variance (MV) and specular receive beamforming schemes in the context of ultrasound imaging of specular reflectors. The main contributions of the study are, 1) a performance comparison of the four receive beamforming schemes through experimental studies for varying angulations of planar reflectors and for reflectors located at varying depths in the medium and, 2) an investigation on the influence of the sub-array length in {MV} beamforming on the imaging of specular structures. The qualitative conclusions are quantitatively validated in terms of contrast and generalized contrast-to-noise ratios. The study examines the benefits and drawbacks of each receive beamforming technique and highlights the significance of application-tailored beamforming schemes for imaging specular structures.
\end{abstract}


\section{Introduction}
Ultrasound (US) imaging typically employs the pulse-echo approach of imaging where the backscattered signals that are received by the transducer are beamformed to reconstruct the final image. The popular receive beamforming schemes in US imaging include the delay and sum (DAS) \cite{Perrot2021SoBeamforming}, filtered delay multiply and sum (DMAS)\cite{Matrone2015TheImaging}, and the minimum variance (MV) beamforming \cite{Synnevag2007AdaptiveImaging} schemes. However,  these receive beamforming schemes assume the medium to be homogeneous and diffuse which gets violated in the presence of specular reflections. The specular reflections occur from structures of dimensions much greater than the ultrasonic wavelength and have high  intensity and directivity that is governed by the angle of incidence of the ultrasound wavefront and the reflector orientation. A specular beamforming (SB) scheme dedicated to imaging of specular structures has also been proposed in literature \cite{Rodriguez-Molares2017SpecularBeamforming}. Contrary to DAS, {DMAS} and {MV}, {SB} utilizes the principles of Snell's law for beamforming and specifically enhances the visualization of the specular reflectors. Nevertheless, there is hardly a comparative study to examine the performance of receive beamforming schemes in the context of imaging specular structures. 

This paper aims to address this gap by comparing and contrasting the performance of {DAS}, {DMAS}, {MV} and, {SB} schemes for specular imaging. The study investigates specular reflectors at varying angulations and depths and also examines the influence of sub-array length ($L_s$) in the {MV} approach for specular imaging. The paper is structured as follows. Section \ref{sec_methods} describes the four beamforming schemes and the experimental setup. The results are presented in Section \ref{sec_Results} and the associated discussion in Section \ref{sec_Disc}. 

\section{Methods} \label{sec_methods}

\subsection{Delay and Sum Beamforming (DAS)}
In {DAS}, the delay compensated signals from all the transducer elements are apodized and summed to form the final beamformed signal, $\mathit{y_{DAS}}$, after coherent compounding across the $\mathit{T}$ transmissions with uniform angular apodization as \eqref{eqn_LitReview_DAS},
\begin{eqnarray}\label{eqn_LitReview_DAS}
y_{DAS}(P) =\sum_{j=1}^{T} \sum_{i=1}^{N_{c}} W_{i}(P)s_{i}(\tau_{P}(i,j))
\end{eqnarray}      
where, $s_{i}(\tau_{P}(i,j)$ represents the delay compensated signal, $\mathit{W_{i}(P)}$ is the apodization weight of the $\mathit{i^{th}}$ transducer element which is proportional to its geometric distance to the pixel $\mathit{P}$. This is because the conventional {DAS} assumes a diffusive homogeneous medium and is independent of the received data \cite{Perrot2021SoBeamforming}. Apodization functions such as rectangular, Hamming, Hanning, Tukey, or Gaussian weighted windows are typically used in {DAS} where, the center of the window is decided by defining an accurate $\mathit{F-number}$ for the transducer considering the directivity of elements \cite{Perrot2021SoBeamforming}.

\subsection{Delay Multiply and Sum Beamforming (DMAS)}
The {DMAS} is a non-linear beamforming algorithm that has been introduced to overcome some of the limitations introduced by {DAS} \cite{Matrone2015TheImaging} by leveraging the coherence in the received data. The delay compensated signals are initially pairwise multiplied in all possible combinations (excluding self-product terms) before summation to form the beamformed signal $\mathit{y_{DMAS}}$ expressed as \eqref{eqn_LitReview_DMAS},
\begin{eqnarray}\label{eqn_LitReview_DMAS}
  y_{DMAS}(P)= \sum_{j=1}^{T}\sum_{i=1}^{N_{c}-1}\sum_{k=i+1} ^{N_{c}} s_{i}'s_{k}'
\end{eqnarray}
where,
\begin{eqnarray}
s_{i}'=&sign(s_{i})\sqrt{|s_{i}(\tau_{P}(i,j))} \label{eqn_LitReview_DMAS_si} \\
s_{k}'=&sign(s_{k})\sqrt{|s_{k}(\tau_{P}(k,j))|} \label{eqn_LitReview_DMAS_sk} 
\end{eqnarray}
where, $s_{i}$, $s_{k}$ are the delay compensated signals from the $\mathit{i^{th}}$ and $(\mathit{i}+1)^{\mathit{th}}$ array elements respectively. The multiplication of signals with similar frequency content synthetically generates a baseband and second harmonic in the spectrum, from which the second harmonics are extracted and filtered with a bandpass filter to obtain the filtered {DMAS} beamformed signal to improve lateral resolution and contrast.

\subsection{Minimum Variance (MV) Beamforming}
In {MV} beamforming, the aperture weights are data-dependent and are adaptively estimated from the signals received in the transducer elements. It aims to improve the resolution through the suppression of off-axis signals and allowing the side lobes in the directions with minimum energy \cite{Synnevag2007AdaptiveImaging}. The beamformed signal ($\mathit{y_{MV}}$) is expressed in vector form with a time-dependent complex apodization weight $\mathit{W}$ for each transmission $\mathit{j}$ as \eqref{eqn_LitReview_MV_No_compound},
\begin{eqnarray} \label{eqn_LitReview_MV_No_compound}
y_{MV}(P,j) = W(P)^Hs(P,j)
\end{eqnarray} 
The receive beamforming with {MV} is viewed as an optimization problem that minimizes the variance or the power of $\mathit{y_{MV}}$ while imposing unit gain in the focal point or the intended direction (distortionless response) and treating all the other signals as interference or noise \cite{Synnevag2007AdaptiveImaging}. The optimization problem is formulated as \eqref{eqn_LitReview_MV_Weight_Optimization}, where $\mathit{a}$ refers to the steering vector, which is a vector of ones for the delay compensated aperture data,
\begin{eqnarray}
\underset{W(P)}{min}{(W(P)^HR(P)W(P)))} \hspace{0.3cm} \textnormal{subject to  } W(P)^Ha = 1
\label{eqn_LitReview_MV_Weight_Optimization}
\end{eqnarray}
\begin{eqnarray} 
R(P) = & E[s(P)s(P)^H]  \label{eqn_LitReview_MV_Cov_Matrix}
\end{eqnarray}
where, $\mathit{E[.]}$ denotes the expectation operator and $\mathit{R(P)}$ represents the spatial covariance matrix estimated as \eqref{eqn_LitReview_MV_Cov_Matrix}. The analytical solution to \eqref{eqn_LitReview_MV_Weight_Optimization} is the adaptive weight vector \eqref{eqn_LitReview_MV_Weights} that is applied to obtain the beamformed signal in \eqref{eqn_LitReview_MV_No_compound}.
\begin{eqnarray}
W(P) = \frac{R(P)^{-1}a}{a^HR(P)^{-1}a}
\label{eqn_LitReview_MV_Weights}
\end{eqnarray}
\subsection{Specular Beamforming}
The specular beamforming {SB} \cite{Rodriguez-Molares2017SpecularBeamforming} is a dedicated beamforming to enhancing specular reflections. It leverages the adherence of specular reflections to Snell’s law to develop a specular transform that maximizes the detection of specular patterns in the received data. The specular beamformed signal (termed as specular transform in \cite{Rodriguez-Molares2017SpecularBeamforming}), $\mathit{y_{SB}}$ is represented as \eqref{eqn_LitReview_SB}, where the propagation delay is a function of transmit ($\alpha_j$) and receive ($\alpha_r$) (and thereby on the reflector orientation $\alpha_g$) angles. 
\begin{eqnarray} \label{eqn_LitReview_SB}
 y_{SB}(P, \alpha_g) = \sum_{j=1}^{T} s(\tau_P(\alpha_j,\alpha_j-2\alpha_g))
\end{eqnarray}  
The work also proposes to enhance the signal-to-noise ratio and the reflection detection probabilities further by correlating the specular transformed signals with a matched filter model, derived to maximize the specular energy. Nevertheless, the method suppresses diffuse reflections and therefore needs to be supplemented with other beamforming techniques. 
\subsection{Experimental Setup}\label{expt_setup}
 Three datasets are acquired with Verasonics Vantage US research platform using L11-5v transducer for 73 plane-wave(PW) transmissions in the range [-18$\degree$, +18$\degree$]. The first dataset is obtained by inserting a 15G needle  into the gel phantom for needle orientations $\alpha_g$ =  $20\degree$, and $30\degree$. Furthermore, a dataset is acquired by inserting a 15G stainless steel needle parallel to the transducer at a depth of 4.3 mm in the uniform gel medium. Another dataset is acquired by inserting the same needle at a depth of 17.5 mm in the same gel medium. All the datasets are later added with speckle noise. The imaging performance is compared against DAS using contrast ratio (CR), and generalized contrast-to-noise ratio (gCNR). Furthermore, two additional \textit{in-vitro} datasets are acquired to compare the beamforming schemes for imaging specular reflectors located at different depths. 
 
The beamformed images are reconstructed from the {RF} datasets using MATLAB R2019a (The MathWorks, Natick, MA, USA) through coherent PW compounding. The {MV} beamforming is performed with $\mathit{L_s}$ = 32, for a temporal sample size of size $5$, and a diagonal loading parameter of $1/10^8$. Further, a comparative study with {MV} beamforming of $\mathit{L_s}$ = 8 and 64 is also performed. The images are shown for a dynamic range of 60 dB.
\section{Results} \label{sec_Results}
\begin{figure}[!t]
\centering
\includegraphics[width=\columnwidth]{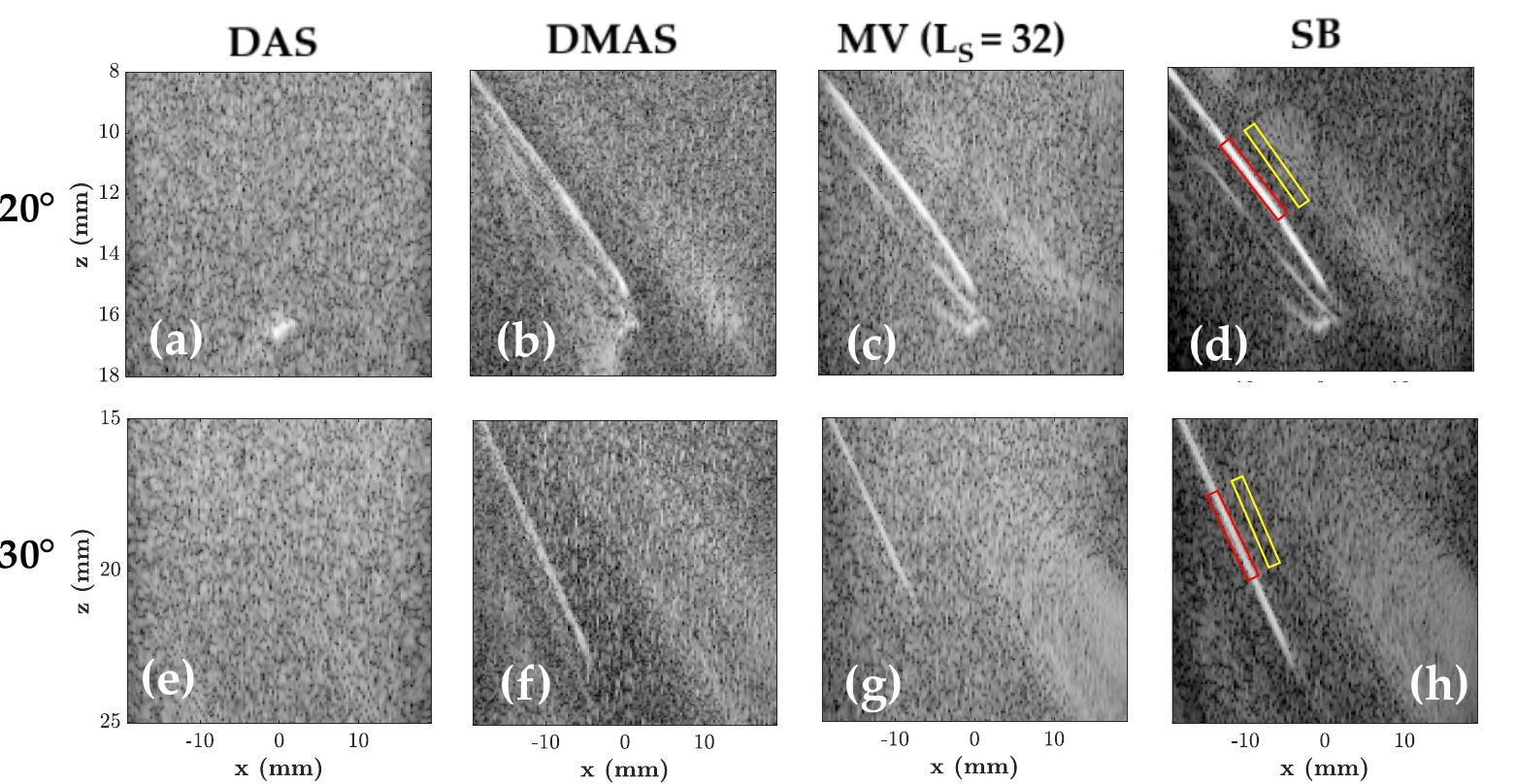}%
 \caption{Beamformed images of the needle inserted into the gel phantom at $20\degree$ (a) {DAS} (b) {DMAS}  (c) {MV} and (d) {SB}. At $30\degree$ (e) {DAS} (f) {DMAS}  (g) {MV} and (h) {SB}.}
\label{fig:RxStudy_Expt_Needle_Speckle_RxBF}
\end{figure}

\begin{figure}[!t]
\centering
\includegraphics[width= \columnwidth]{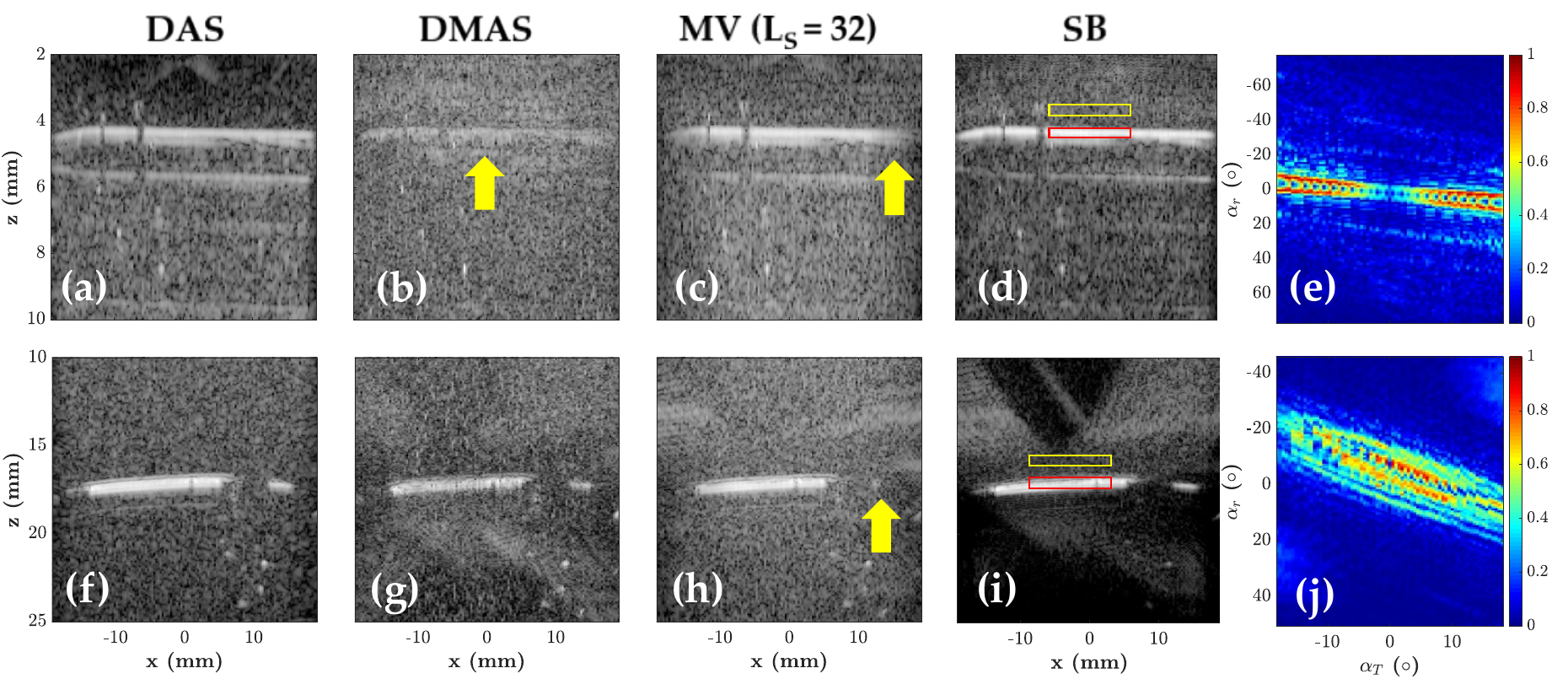}%
\caption{Comparison of the beamforming schemes for needle at a depth of 4.3 mm (a) {DAS} (b) {DMAS} (c) {MV} (d) {SB}. (e) Distribution of the reflected energy from a pixel located at $\mathit{(x,z)}$ = (0.45, 4.3) mm. For needle at a depth of 17.5 mm  (f) {DAS} (g) {DMAS} (h) {MV} (i) {SB} and (j) Distribution of the reflected energy from a pixel located at $\mathit{(x,z)}$ = (-1.6, 17.5) mm.}
\label{fig:RxStudy_Expt_Needle_DepthStudy}
\end{figure}

Row 1 in Fig. \ref{fig:RxStudy_Expt_Needle_Speckle_RxBF} are the beamformed images for   $\alpha_g$ = $20\degree$, and row 2 for $\alpha_g$ = $30\degree$ with the {DAS}, filtered {DMAS}, {MV}, and {SB} schemes respectively arranged along columns 1, 2, 3, and 4. The reflector is better visualized in {PW} images with filtered {DMAS}, {MV}, and {SB} when compared to the {PW}-{DAS} images  where the reflector is significantly degraded by the speckle noise. The quantitative metrics {CR}, and {gCNR}, are presented in the Table \ref{tab:RxStudy_metrics} to confirm the qualitative conclusions. 
\begin{table}[]
\caption{Comparison of {CR}, and {gCNR}}
\label{tab:RxStudy_metrics}
\centering
\begin{tabular}{cclcl}
\hline
\hline
\multirow{2}{*}{$\boldsymbol{\alpha_g}$ \textbf{($\degree$)}} & \multicolumn{4}{c}{\textbf{DAS/DMAS/MV/SB}}            \\ \cline{2-5} 
& \multicolumn{2}{c}{\textbf{CR(dB)}} & \multicolumn{2}{c}{\textbf{gCNR}}                \\ \hline
20 & \multicolumn{2}{c}{-0.80/13.61/{18.46}/$\boldsymbol{26.8}$} & \multicolumn{2}{c}{0.47/0.90/$\boldsymbol{0.99}$/$\boldsymbol{0.98}$} \\ 
30 & \multicolumn{2}{c}{-3.40/9.61/{8.7}/$\boldsymbol{19.8}$}    & \multicolumn{2}{c}{0.51/0.68/0.78/$\boldsymbol{0.92}$} \\
\hline
\begin{tabular}[c]{@{}c@{}}0\\ (4.3 mm)\end{tabular} &
\multicolumn{2}{c}{27.92/7.11/24.38/$\boldsymbol{29.68}$} &
 \multicolumn{2}{c}{$\boldsymbol{1}$/0.79/0.98/0.98} \\
\begin{tabular}[c]{@{}c@{}}0\\ (17.5 mm)\end{tabular} &
  \multicolumn{2}{c}{30.62/24.89/24.97/$\boldsymbol{33.61}$} &
  \multicolumn{2}{c}{$\boldsymbol{1}$/$\boldsymbol{1}$/$\boldsymbol{1}$/$\boldsymbol{1}$} \\
\hline
\hline
\end{tabular}
\end{table}
The beamforming schemes are compared for their performance in imaging specular reflectors at two different depths in Fig. \ref{fig:RxStudy_Expt_Needle_DepthStudy}. Fig. \ref{fig:RxStudy_Expt_Needle_DepthStudy}(a)-(d) show respectively the {DAS}, filtered {DMAS}, {MV}, and {SB} images for the needle at a depth of 4.3 mm in the gel phantom. The performance of filtered {DMAS} is severely degraded as the reflector is hardly visible when compared to the other beamforming schemes. The spatial distribution of the reflected energy from a pixel located on the specular reflector is plotted in Fig. \ref{fig:RxStudy_Expt_Needle_DepthStudy}(e). The $\mathit{x}$ axis indicates the {PW} transmissions between [-$18\degree$, +$18\degree$], and the $\mathit{y}$ axis is the range of angles ($\alpha_r$) made by the pixel with each of the $\mathit{N_c}$ transducer elements. The plot shows that only a few elements that are closer to the pixel of interest ($\alpha_r$ =  $0\degree$) are receiving the reflected energy across the {PW} transmissions. It is also noted that {DAS} and {SB} images in Fig. \ref{fig:RxStudy_Expt_Needle_DepthStudy}(a) and (d) resolve the reflector better towards the edges of the scan grid (between $\mathit{x}$ = -15 mm to -10 mm and $\mathit{x}$ = 10 mm to 15 mm). However, when the reflector is at a greater depth of 17.5 mm, {DMAS} in Fig. \ref{fig:RxStudy_Expt_Needle_DepthStudy}(g) is of comparable visual quality to {DAS}, {MV} and {SB} images in  Fig. \ref{fig:RxStudy_Expt_Needle_DepthStudy}(f), (h) and, (i) respectively. Contrary to the previous case, the plot of the spatial distribution of reflected energy from the pixel on the reflector in (Fig. \ref{fig:RxStudy_Expt_Needle_DepthStudy}(j)), shows that more transducer elements receive the reflections. A small discontinuity exists in the reflector between $\mathit{x}$ = 5 mm - 12 mm which is due to the air content in the gel medium during the experiment and is captured in the {DAS}, {DMAS}, and {SB} images, but is not very evident in the {MV} image (Fig. \ref{fig:RxStudy_Expt_Needle_DepthStudy}(h)). The quantitative metrics for the two cases are presented in Table \ref{tab:RxStudy_metrics} (last two rows). The specular {ROI} is selected on the needle and the background region is selected just outside the needle as annotated in the {SB} images in Fig. \ref{fig:RxStudy_Expt_Needle_DepthStudy} in red and yellow respectively. Interestingly, in both cases, the {DAS} shows comparable metric values with {SB} and higher metric values consistently over {DMAS} and {MV} beamforming.
\begin{figure}[!h]
\centering
\includegraphics[width=\columnwidth]{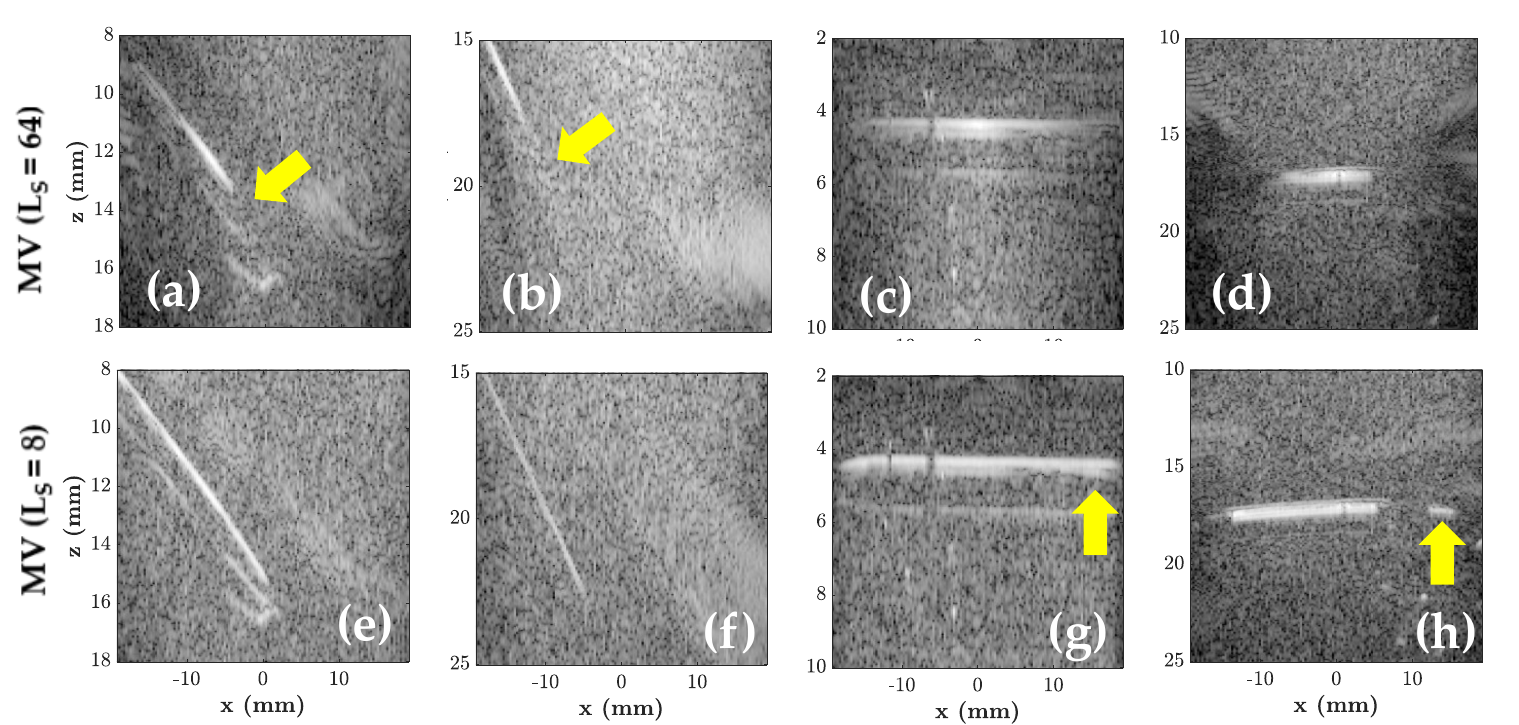}%
  \caption{{MV} beamformed images with $\mathit{L_s}$ = 64 for the needle inserted into the gel phantom at  (a) $20\degree$ and (b) $30\degree$. For the needle at 0$\degree$ at a depth of (c) 4.3 mm and (d) 17.5 mm. {MV} beamformed images with $\mathit{L_s}$ = 8 at (e) $20\degree$ and (f) $30\degree$. For the needle at 0$\degree$ at a depth of (g) 4.3 mm and (h) 17.5 mm.}
\label{fig:RxStudy_Expt_Needle_MVL64}
\end{figure}

Finally, the influence of $\mathit{L_s}$ in {MV} beamforming on the imaging of specular reflectors are analysed and presented in Fig. \ref{fig:RxStudy_Expt_Needle_MVL64}. The {MV} beamformed images reconstructed with  $\mathit{L_s}$ = 64 in row-1 and with  $\mathit{L_s}$ = 8 in row-2 of Fig. \ref{fig:RxStudy_Expt_Needle_MVL64}. It is seen that the performance of {MV} in imaging the specular reflectors is adversely affected by  $\mathit{L_s}$ = 64 when compared to the corresponding images for $\mathit{L_s}$ = 32 and 8. The needle appears distorted with only a small section of the structure visible in the beamformed images. For $\mathit{L_s}$ = 8, the needle structures are enhanced particularly at the entry point and the tip of the needles when compared to $\mathit{L_s}$ = 32 in Fig. \ref{fig:RxStudy_Expt_Needle_Speckle_RxBF}. Further, in Fig. \ref{fig:RxStudy_Expt_Needle_MVL64}(g), it is seen that the reflector is better resolved towards the edges of the scan grid (between $\mathit{x}$ = -15 mm to -10 mm and $\mathit{x}$ = 10 mm to 15 mm). Similarly, the small discontinuity in the reflector which is not very evident in Fig. \ref{fig:RxStudy_Expt_Needle_DepthStudy}(h) is captured with $\mathit{L_s}$ = 8 in Fig. \ref{fig:RxStudy_Expt_Needle_MVL64}(h). 

\section{Discussion}\label{sec_Disc}
This paper compared the {DAS}, {DMAS}, {MV}, and specular beamforming techniques in the imaging of planar specular structures for {PW} transmissions. The study focused on validating the receive beamforming schemes for various orientations and depths of the structures relative to the transducer array. Additionally, the study evaluated the significance of choosing $\mathit{L_s}$ appropriately for imaging specular surfaces with {MV} beamforming.  The results indicate that the {DMAS}, {MV}, and {SB} techniques outperform {DAS} beamforming in imaging specular reflectors at different reflector angles. This is because the {DMAS} and {MV} beamforming techniques do not rely on a data-independent/geometry-driven apodization like {DAS} whereas {SB} leverages the physics of specular reflections to exclusively image the specular reflectors.

In {DMAS} beamforming, the apodization is inherent from the pair-wise multiplication of signals received by the transducer elements \cite{Matrone2015TheImaging}. The multiplication is effectively an autocorrelation of the receive aperture in the reflections received across the transducer array elements. As a result, the pair-wise multiplication emphasizes the contribution of specular reflections due to coherence enhancement by auto-correlation in the beamformed signal, even when reflections from inclined reflectors are reflected off-axis based on the angle of incidence of the {PW} transmission. When comparing different beamforming schemes for imaging specular reflectors at varying depths, {DMAS} underperforms {DAS}, {MV} beamforming, and {SB}, especially at shallower depths closer to the transducer (as shown in Fig. \ref{fig:RxStudy_Expt_Needle_DepthStudy}(a)-(d)). This is because reflections are received only by the transducer elements closest to the reflector pixels due to the limited range of reflection angles at these depths, as illustrated in Fig. \ref{fig:RxStudy_Expt_Needle_DepthStudy}(e). The inherent auto-correlation of the aperture in {DMAS}, which is expected to improve coherence, contributes to clutter due to minimal reflections available in the farther transducer elements. Consequently, the geometry-driven apodization in {DAS} is ideal and is a suitable candidate for shallow-depth applications. However, when the reflector is at a greater depth (Fig. \ref{fig:RxStudy_Expt_Needle_DepthStudy}(f)-(i)), the performance of {DMAS} is improved as the reflected energy is available in more transducer elements as seen in Fig. \ref{fig:RxStudy_Expt_Needle_DepthStudy}(j). 

The {MV} beamforming with $\mathit{L_s}$ = 32 gives a consistent performance in all cases when compared to {DAS} and {DMAS}. However, a slight deterioration of the image quality of the reflectors is observed at the edges of the scan grid in Fig. \ref{fig:RxStudy_Expt_Needle_DepthStudy}. This is because the {MV} approach assumes sufficient signal-to-noise ratio in the subarrays which may be minimal from the pixels closer to the edges of the scan plane. A smaller sub-array length is more optimal towards the edges of the scan grid. The performance of {MV} beamforming while imaging specular structures is adversely affected when $\mathit{L_s}$ is not selected correctly. Typical adaptive beamforming schemes like {MV} seek to minimize the variance in the received data to determine the best set of receive aperture weights to improve contrast and resolution. The optimal $\mathit{L_s}$ for the minimization problem is proposed as $\mathit{L_s} \leq{N_c/2}$. A smaller $\mathit{L_s}\in[1, N_c/2]$  improves the robustness by trading off the resolution and higher $\mathit{L_s}$ improves the resolution by compromising robustness.  Nevertheless, the results presented in Fig. \ref{fig:RxStudy_Expt_Needle_MVL64} illustrated that an enhanced visualization of the specular structures is subjected to choosing correct $\mathit{L_s}\in[1, N_c/2]$. A $\mathit{L_s}$ = 64 worsens and a $\mathit{L_s}$ = 8 enhances the output of {MV} in the context of specular imaging. This is because the reflected energy from the specular reflector is concentrated on a few transducer elements, unlike the diffuse scattering where the scattered signals are spatial distributed across the  transducer array. It is found that the reflected specular energy approximately is distributed within 32 transducer elements, therefore a $\mathit{L_s}>$32 accumulates noise during the estimation of spatial covariance. However, the selection of optimal $\mathit{L_s}$ for enhancing specular reflectors should not compromise the resolution and contrast of the soft tissue structures. A smaller $\mathit{L_s}$ as 8 degrades the resolution of soft tissue structures when compared to $\mathit{L_s}$ = 64. Therefore, an optimization framework needs to be developed to tune $\mathit{L_s}$ according to the imaging scenario. This can add to the existing computational complexity of {MV}. 

Upon analyzing the performance of {SB}, it is found to be superior to all other beamforming techniques. However, as anticipated, the improvement is focused on specular structures, resulting in the suppression of soft tissue structures (i.e., non-specular structures). This is not ideal for real-time imaging, as the suppression of such tissue structures can lead to the loss of crucial tissue information (such as missing important anatomical landmarks) and create diagnostic challenges.

It is emphasized that all the discussed receive beamforming techniques have their unique advantages, but also have limitations while imaging specular interfaces. Hence, it is not feasible to directly apply any of these techniques for imaging specular structures without considering the application. 

\section*{Acknowledgment}
The authors would like to acknowledge the facilities of the Center for Computational Imaging, Indian Institute of Technology Palakkad (IITPKD), India, and Mr. Ananthu Sasikumar, Junior Technician, IITPKD  for his assistance in logistics.


\bibliographystyle{IEEEtran}
\bibliography{references}
\
\vfill
\end{document}